\documentclass[useAMS,usenatbib]{mn2e}
\usepackage{amsmath}
\usepackage{amssymb}
\usepackage{graphicx}
\usepackage{bm}
\title{Impact of dimensionless
numbers on the efficiency of MRI-induced turbulent transport.}
\author[Lesur, G. and P-Y. Longaretti]{Lesur, G.$^1$ and P-Y. Longaretti$^1$\\
$^1$ Laboratoire d'Astrophysique Observatoire de Grenoble\\
BP 53, 38041 GRENOBLE CEDEX 9 }

\begin{document}

\date{Accepted ????. Received ?????; in original form ???}

\pagerange{\pageref{firstpage}--\pageref{lastpage}} \pubyear{2007}

\maketitle

\label{firstpage}
\begin{abstract}
The magneto-rotational instability is presently the most promising
source of turbulent transport in accretion disks. However, some
important issues still need to be addressed to quantify the role
of MRI in disks; in particular no systematic investigation of the
role of the physical dimensionless parameters of the problem on
the dimensionless transport has been undertaken yet. For
completeness, we first generalize existing results on the marginal
stability limit in presence of both viscous and resistive
dissipation, exhibit simple scalings for all relevant limits, and
give them a physical interpretation. We then reexamine the
question of transport efficiency through numerical simulations in
the simplest setting of a local, unstratified shearing box, with
the help of a pseudo spectral incompressible 3D code; viscosity
and resistivity are explicitly accounted for. We focus on the
effect of the dimensionless magnetic field strength, the Reynolds
number, and the magnetic Prandtl number. First, we complete
existing investigations on the field strength dependence by
showing that the transport in high magnetic pressure disks close
to marginal stability is highly time-dependent and surprisingly
efficient. Second, we bring to light a significant dependence of
the global transport on the magnetic Prandtl number, with
$\alpha\propto Pm^\delta$ for the explored range: $0.12<Pm<8$ and
$200<Re<6400$ ($\delta$ being in the range 0.25 to 0.5). We show
that the dimensionless transport is not correlated to the
dimensionless linear growth rate, contrarily to a largely held
expectation. For large enough Reynolds numbers, one would expect
the reported Prandtl number scaling of the transport should
saturate, but such a saturation is out of reach of the present
generation of supercomputers. Understanding this saturation
process is nevertheless quite critical to accretion disk transport
theory, as the magnetic Prandtl number $Pm$ is expected to vary by
many orders of magnitude between the various classes of disks,
from $Pm \ll 1$ in YSO disks to $Pm\gtrsim$ or $\gg 1$ in AGN
disks. More generally, these results stress the need to control
dissipation processes in astrophysical simulations.
\end{abstract}
%
%%-----------------------------
%%      your text
%%-----------------------------

\section{Introduction}

Angular momentum transport has always been a central issue in
accretion disk theory. The first $\alpha$ model \citep{SS73}
already assumed the presence of strong turbulent motions, modelled
through an effective viscosity, orders of magnitude larger than
the expected disk molecular viscosity. Since then, the physical
origin of this turbulence has been highly debated. As purely
hydrodynamic non stratified Keplerian flows are known to be
linearly stable to small perturbations, a finite amplitude
instability was first envisioned to trigger turbulence. This
question was studied both experimentally \citep{RZ99,R01,RDDZ01}
and numerically \citep{BHS96,HBW99}, leading to a long
controversy. More recent numerical and experimental investigations
of this problem strongly support the idea that the transport due
to this mechanism, if present, would be far to inefficient to
account for the short disk evolution time-scales imposed by
astrophysical observations \citep{LL05,JBSG06}. Linear
instabilities of hydrodynamic origin have also been envisioned as
a source of turbulence, relating in particular to the flow
stratification \citep{KB03,U03,DMNRHZ05,SR05}, but these are
either not present or too inefficient (\citealt{JG06,AU04,LL07};
see \citealt{LL05} and references therein for a recent review of
this issue).

The potential role of MHD instabilities in accretion disks was
pointed out in a seminal paper by \cite{BH91}, devoted to an
analysis of what is now known as the magneto-rotational
instability (MRI). This instability has extensively been studied
since then, mainly with the help of local \citep{HGB95,SHGB96} and
global \citep{H00} 3D numerical simulations. Although a more
recent set of numerical simulations did focus on MRI energetics
\citep{GS05}, the dissipation of turbulent fields in these
simulations is not controlled, as no physical term was introduced
to account for physical viscosity and resistivity. Note however
that \cite{BNST95} have introduced such dissipation in their
simulations, but kept it as small as possible, and in any case did
not try to investigate their effect in a systematic way.
Resistivity effects alone have also been introduced by
\cite{FSH00}, but viscous effects were still neglected. This
raises questions about the exact role of numerical dissipation in
all these simulations, especially at the light of our recent
investigation of subcritical turbulence in accretion disks
\citep{LL05}, which clearly showed that a careful control of
dissipation and resolution --- and more generally of the
dimensionless parameters of the problem --- is required to
properly quantify turbulent transport.

This issue is addressed here in the context of MRI-driven
turbulence, using a 3D spectral Fourier code, which allows a
precise monitoring of viscous, resistive and numerical
dissipation. First, we recall the MHD equations in the shearing
sheet framework \citep{HGB95}, along with the relevant
dimensionless parameters of the problem, and summarize what is
known about their effect on MRI-induced turbulent transport. Next,
we investigate the linear stability of the MRI, which to the best
of our knowledge has not been fully characterized when both
viscosity and resistivity are accounted for in the dispersion
relation. Then, we present new results on the behavior of
turbulent transport in dimensionless parameters regime that have
not been investigated in the past: first, very close to the
threshold of instability (in terms of relative magnetic field
strength), and then with respect to the magnetic Prandtl number,
which has been ignored in all previous investigations. The
dependence of turbulence transport on the magnetic Prandtl number
is the most significant finding of this investigation. This
dependence may turn out to be critical, as the magnetic Prandtl
number varies by many order of magnitudes in astrophysical disks.
The astrophysical implications of our findings are further
discussed in our concluding section, along with potential caveats
relating to numerical limitations that may influence our results.

\section{Shearing box characterization and summary of earlier results\label{shearbox}}

The MRI has already been extensively studied in the literature
(see, e.g., \citealt{B03} and references therein for a review of
the subject). Our objective is to extend previous work through a
systematic exploration of the dependence of the MRI-induced
transport on the physical quantities characterizing the problem.
For simplicity, we work in a shearing sheet setting (see
\citealt{HGB95} for a description of the shearing box equations,
numerical boundary conditions, and conserved quantities); vertical
stratification is ignored, but both viscous and resistive
microphysical (molecular) dissipation are included. This differs
from previous investigations, where this is always ignored. Our
previous experience with subcritical hydrodynamic transport has
shown us that the inclusion of explicit dissipation is required to
precisely characterize transport properties and ro sort out
converged simulations from under-resolved ones (see \citealt{LL05}
for an extensive discussion and illustration of these points).

The problem is formulated in a cartesian frame centered at
$r=R_0$, rotating with the disk at $\Omega=\Omega(R_0)$ with
radial dimension $H\ll R_0$. In this work, $H$ is the size of our
simulation boxes, in the vertical and radial dimensions. This
leads to the following set of equation, assuming $\phi\to x$,
$r\to -y$:

\begin{eqnarray}
\label{base1}\partial_t \bm{u}+\bm{u}\cdot \bm{\nabla}
\bm{u}&=&-\frac{1}{\rho}\bm{\nabla}P+\frac{1}{\mu_0
\rho}(\bm{\nabla}\times \bm{B})\times\bm{B},\\
\nonumber
& & -2\bm{\Omega}\times\bm{u}-2\Omega Sy\bm{e_y}+\nu\bm{\Delta}\bm{u}\\
\label{base2}\partial_t \bm{B}&=&\bm{\nabla}\times(\bm{u}\times\bm{B})+\eta\bm{\Delta{B}},\\
\label{base3}\nabla \cdot \bm{u}&=&0,\\
\label{base4}\nabla\cdot\bm{B}&=&0,
\end{eqnarray}

\noindent where the medium is defined by $S=-r\partial_r\Omega$.
For simplicity, incompressible motions are assumed. This is a
priori justified by the fact that MRI-induced motions are usually
subsonic, so that one expects at least in first approximation that
compressibility effects do not play a major role in the problem.
This approximation allows us to remove the flow Mach number from
the list of dimensionless parameters characterizing the problem,
so that we can more effectively isolate and quantify the role of
the various physical agents.

The terms on the right-hand side member of Eq.~(\ref{base1}) are
the gas pressure, Lorentz force, Coriolis force, tidal force, and
viscous dissipation, respectively. The steady-state solution to
this equation is the local profile $\bm{u}=Sy\bm{e_x}$ with
$S=3/2\Omega$ for Keplerian disks. In the remainder of this paper,
we will use the deviation from the laminar profile $\bm{w}$
defined by $\bm{w}=\bm{u}-Sy\bm{e_x}$. For simplicity, we assume
that the steady-state magnetic field $B_0$ lies along the vertical
axis. Note that this field is also the average field in the
shearing sheet box, and is conserved during the evolution thanks
to the shearing sheet boundary conditions \citep{HGB95}.

These equations are characterized by four dimensionless numbers,
the first three relating to the Navier-Stokes equation while the
last one belongs to the induction equation:

\begin{itemize}
  \item The Reynolds number, $Re\equiv SH^2/\nu$, measuring the
  relative importance of nonlinear coupling through the advection
  term, and viscous dissipation.
  \item A proxy to the plasma beta parameter, defined here as $\beta=S^2
  H^2/V_A^2$ where $V_A^2= B_o^2/\mu_o \rho$ is the Alfv\'en speed. The
  rationale of this definition follows from the vertical
  hydrostatic equilibrium constraint $c_s \sim \Omega H$, which is
  expected to hold in thin disks, so that our definition of
  $\beta$ is indeed related to the plasma parameter in an
  equivalent, vertically stratified disk. This parameter measures
  the relative weight of the Lorentz force and the advection term.
  \item The rotation number (inverse Rossby number), defined as
  $R_\Omega = 2\Omega/S$, which measures the relative importance
  of the Coriolis force.
  \item The magnetic Reynolds number, $R_m=SH^2/\eta$, which
  measures the relative importance of resistive dissipation with respect
  to the ideal term in the induction equation.
\end{itemize}

We consider only Keplerian disks in this investigation, so that
the rotation number is held fixed to its Keplerian value
$R_\Omega=-4/3$. This leaves us with three independent
dimensionless numbers: $\beta$, $Re$, and $Rm$.

On the other hand, the (local in the disk) dimensionless transport
coefficient,

\begin{equation}\label{alpha-def}
  \alpha=\frac{\langle v_x v_y - B_x B_y/(\mu_o\rho)\rangle}{S^2H^2},
\end{equation}

\noindent being a dimensionless number, can only depend on the
local dimensionless parameters characterizing the flow that we
have just defined\footnote{It may also depend on the simulation
aspect ratio and resolution, from a numerical point of view.} (the
bracketing refers to appropriate box and/or time averages). Our
task reduces to characterize this dimensionless transport, as a
function of the three independent dimensionless numbers just
defined. However, for later convenience, we take them to be
$\beta$, $Re$ and $P_m\equiv \nu/\eta=Rm/Re$ instead (the
rationale of this latter choice will become apparent later on).

All previous investigations ignore the dependence on the last two
dimensionless numbers, who have not been included in the physical
description up to now. Within such an approximation, \citet{HGB95}
have characterized the dependence of $\alpha$ on $\beta$. Their
results imply that

\begin{equation}\label{scaling}
\alpha\simeq 3 \beta^{-1/2},
\end{equation}

\noindent from their Eqs.~(10), (15), (16) and (18).

This implies in particular that $\alpha$ increases when the
initial (and box average) magnetic field $B_o$ is increased.
However, for a large enough field, the smallest unstable
wavelength (which increases along with $B_o$) becomes larger than
the box size, and the instability is quenched. On this basis, one
expects that the scaling Eq.~(\ref{scaling}) would break down
close enough to the critical $\beta$ stability limit. This
question is somewhat investigated in the present work. However,
most of our effort is devoted to characterizing the $Re$ and $Pm$
dependence of $\alpha$.

\section{Linear stability analysis}\label{analytic}

In order to quantify the MRI induced turbulent transport, it is
first necessary to define the parameter domain in which this
instability operates. The linear stability of differentially
rotating disks in presence of a magnetic field was first
investigated in the astrophysical context by \cite{BH91}. Then,
the instability in the weakly ionized case has been considered
\citep{BB94,W99,BT01}, leading to the well known Dead Zone problem
\citep{G96}. However, we are not aware of any reasonably complete
and heuristically clarified investigation of the stability limits
of the fluid when both viscous and resistive dissipation are taken
into account. Some discussions of this point are available in the
literature, mostly motivated by liquid-metal experiments, in the
limit $Pm\ll 1$ \citep{JGK01,RS02}. However, these papers exhibit
no clear asymptotical limit that may be useful for astrophysical
disks. Therefore, we provide such an analysis here, as a prelude
to our nonlinear simulations.

We will consider only axisymmetric perturbations, so that we can
eliminate the azimuthal perturbation transport term. Note that
this assumption does not seem to have a great influence on the
stability limit, since 3D numerical simulations and linear
analysis of axisymmetric modes exhibit nearly the same stability
limit; this holds in particular in the simulations presented here.

We linearize and Fourier transform the equations of motion by
assuming $\bm{v}=\bm{v_0}\exp\big(i(\omega t-k_y y-k_z z)\big)$
and $\bm{b}=\bm{b_0}\exp\big(i(\omega t-k_y y-k_z z)\big)$. This
yields the following linearized equation set:

\begin{eqnarray}
\label{linv}(i\omega+\nu k^2) \bm{v_0}&=&i\,\bm{k}\psi-i\,k_z \frac{B_0}{\mu_0\rho_0} \bm{b_0}\\
\nonumber                 & &+(2\Omega-S)v_y\bm{e_x}-2\Omega v_x \bm{e_y},\\
\label{linb}(i\omega+\eta k^2) \bm{b_0}&=&-i\,k_zB_0\bm{v}+b_yS\bm{e_x},\\
i\,\bm{k}\cdot \bm{v}&=&0,\\
i\,\bm{k}\cdot \bm{B}&=&0,
\end{eqnarray}

\noindent where $\psi$ is the perturbation in total pressure
$(P+B^2/\mu_0)/\rho$. Introducing $\omega_\nu\equiv\omega-i\nu
k^2$ and $\omega_\eta\equiv\omega-i\eta k^2$, the Alfv\'en speed
$V_A^2=B_0^2/\mu_0\rho$, the epicyclic frequency
$\kappa^2=2\Omega(2\Omega-S)$ and $\gamma^2=k_z^2/k^2$, one
eventually gets the dispersion relation:

\begin{eqnarray}
\nonumber(\omega_\nu \omega_\eta-k_z^2V_A^2)\Bigg(\omega_\nu^2
\omega_\eta^2-2\omega_\nu \omega_\eta k_z^2V_A^2-\omega_\eta^2\kappa^2\gamma^2& & \\
\label{disp}-k_z^2V_A^2\Big(2\Omega
S\gamma^2-k_z^2V_A^2\Big)\Bigg)&=&0
\end{eqnarray}

\noindent which we now solve in various dissipation regimes.

\subsection{$Pm=1$ behavior}

\begin{figure}
      \centering
      \includegraphics[height=5cm]{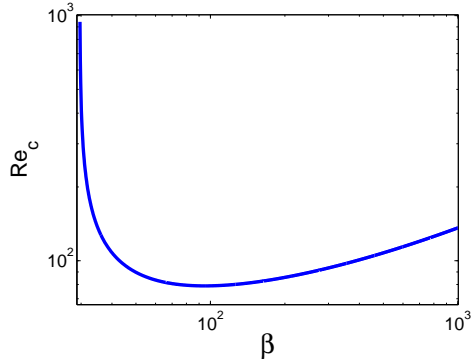}
      \caption{MRI linear stability limit\label{mri_stab} for $Pm=1$}
\end{figure}

Let us first look at the $Pm=1$ case, where the dispersion
equation can be solved exactly by analytical means. The condition
$\Im(\omega)<0$ expresses the existence of the instability, and
implies that the MRI exists if and only if
$\nu^2k^4<-\omega_\nu^2$. From this constraint and the dispersion
relation Eq.~(\ref{disp}), one finds that:

\begin{equation}
\label{nu}
\nu^2<\frac{\sqrt{\kappa^4\gamma^4+16k_z^2V_A^2\Omega^2\gamma^2}}{2k^4}-
\frac{k_z^2V_A^2}{k^4}-\frac{\kappa^2\gamma^2}{2k^4},
\end{equation}

\noindent is a necessary and sufficient criterion for instability.
One can check that the highest $\nu$ values obtain when $\gamma=1$
and $k_z=\mathrm{min}(k_z)=2\pi/H$, which corresponds to the
so-called channel flow solution in the $z$ direction. From our
definition of the Reynolds number as $Re=SH^2/\nu$ where $H$ is
the numerical box height or the typical disk height, and of the
plasma parameter $\beta=S^2H^2/V_A^2$, the stability limit
Eq.~(\ref{nu}) translates into a relation between these two
parameters, represented on Fig.~ \ref{mri_stab}.

Note that the instability has two different limits, depending on
the $\beta$ parameter:
\begin{itemize}
\item A high $\beta$ regime, corresponding to a low magnetic pressure.
In this regime, marginal stability occurs at a characteristic
Reynolds number value $Re_c\simeq 80$. This behavior illustrates
that the growth time scale of the most unstable mode must be
shorter than the dissipation time scale, defined by $\tau_d\simeq
k^2/\nu$.
\item A low $\beta$ regime, which is nearly Reynolds independent.
In this region, one can define a critical $\beta$ ($\beta_c=29.5$)
for which the MRI is lost. This behavior can be explained by
considering the unstable mode of shortest wavelength: as $\beta$
goes to smaller values, the smallest unstable wavelength increases
(see Eq.~\ref{nu}). At some point it becomes larger than the scale
height $H$ (or box size in our case) and the instability is lost.
Since this phenomenon takes place at large scales, the Reynolds
number plays little or no role. Note that this regime is not
specific to our unstratified calculation, since similar results
are found for a stratified medium where marginal stability usually
occurs for $\beta_c \gtrsim 1$ (see e.g \citealt{BH91} and
\citealt{GB94}). This limit is reached when the last factor in
Eq.~(\ref{disp}) cancels out, i.e., when $2\Omega S = V_A^2 k_z^2$
(the usual dissipationless MRI stability limit).
\end{itemize}

\subsection{$Pm \ne 1$ behavior}

The dispersion relation can no longer be solved exactly in this
case, but an approximate solution can be found in the low magnetic
field limit ($V_A\rightarrow 0$, or more precisely $V_Ak_z \ll
\kappa$), where marginal stability follows from a balance between
the destabilizing term, and the dissipation ones. The ``opposite"
(high $\beta$) marginal stability limit, where destabilization is
balanced by the usual Alfv\'enic magnetic tension, is briefly
addressed at the end of this section.

In the limit of vanishing magnetic field, the dispersion relation
has two relevant roots $\omega_\eta^2=0$ and
$\omega_\nu^2=\kappa^2$. In what follows, we refer to these roots
as the Alfv\'enic and the inertial branch, respectively. Looking
for the first order correction in $V_A^2 k_z^2$ to the Alfv\'enic
branch yields the following result, which describes the MRI modes:

\begin{equation}
\label{omega_lowB} \omega=i\eta k_z^2\pm i\Bigg(\frac{2\Omega
S}{k_z^4(\eta-\nu)^2+\kappa^2}\Bigg)^{1/2}V_Ak_z.
\end{equation}

Note that viscosity and resistivity do not play a symmetric role
in this expression. Two interesting limits with respect to the
magnitude of the viscosity prove useful to characterize marginal
stability. As before, we maximize instability by assuming
$\gamma=1$ and $k_z=2\pi/H$.

\subsubsection{Low viscosity limit:}

First consider the limit where $\nu k_z^2 \ll \kappa$. In this
case, Eq.~(\ref{omega_lowB}) reduces to $\eta k^2= (2\Omega
S/\kappa^2)^{1/2}V_A k_z$ (where $\eta k_z^2 \ll \kappa$ has been
self-consistently used). Using the Lundquist number defined as
$Lu=Rm\beta^{-1/2}$, this can be recast as

\begin{equation}\label{Rm}
 Lu=\left(\frac{2\pi}{3^{1/2}}\right)\simeq 3.6.
\end{equation}

\noindent Note that our definition of the Lundquist number is not strictly
identical to \cite{T06} but is widely used in the MHD community\footnote{The
difference lies in the fact that our calculation is made in the limit of high
$\beta$, leading to a linear growth rate controled by $V_A$ instead of $\Omega$}
In this regime, the $\omega_\eta^2\kappa^2$ term
balances the $2\Omega S V_A^2 k_z^2$ term in the dispersion
relation Eq.~(\ref{disp}). Eq.~(\ref{Rm}) corresponds to the limit
found by \cite{FSH00}. It is also related to the origin of the
``dead zone" in accretion disks (see e.g \citealt{G96}). This
marginal stability limit is relevant to disks with low Prandtl
numbers ($Pm \ll 1$) and high Reynolds numbers ($Re \gg 1$), such
as YSO disks.

Also, for negligible resistivity, growth rates in this regime are
given by

\begin{equation}\label{GRRm}
  \tau^{-1}\simeq \frac{1}{2\pi}\left(\frac{2\Omega
  S}{\kappa^2}\right) V_A k_z.
\end{equation}

\noindent This result is valid for $V_A k_z \lesssim \kappa$ due
to our expansion scheme; it also gives the correct order of
magnitude of maximum growth rates when $V_A k_z \sim \kappa$, as
shown by the standard dissipationless MRI analysis.

\subsubsection{High viscosity limit:}

Conversely, consider the large viscosity limit, where $\nu k^2 \gg
\kappa$. The corresponding relations in this limit are

\begin{equation}\label{RmRe}
  Re Rm=\frac{3^{1/2}}{2}(2\pi)^3 \beta^{1/2}\simeq 215 \beta^{1/2}.
\end{equation}

\noindent and

\begin{equation}\label{GRReRm}
  \tau^{-1}\simeq \frac{1}{2\pi}\left(\frac{2\Omega
  S}{\nu k_z^2}\right) V_A k_z.
\end{equation}

\noindent In this regime, the $\omega_\nu^2 \omega_\eta^2$ term
balances the $2\Omega S V_A^2 k_z^2$ term in the dispersion
relation Eq.~(\ref{disp}). The growth rates relevant here are much
smaller than in the small viscosity limit, Eq.~(\ref{GRRm}). In
fact, Eq.~(\ref{omega_lowB}) indicates that this is the case as
soon as $\nu k_z^2 \lesssim \kappa$, or equivalently, for the
largest mode, when

\begin{equation}\label{Relim}
Re \gtrsim 3(2\pi)^2/2\simeq 60.
\end{equation}

\noindent This limit divides the low and high viscosity regime.

The marginal stability limit Eq.(\ref{RmRe}) obtains for large
Prandtl and small Reynolds numbers. In the large Prandtl ($Pm \gg
1$) and large Reynolds number limit ($Re \gg 1$) expected in AGN
disks, the growth rates of Eq.~(\ref{GRRm}), or more generally of
dissipationless MRI, are recovered. As before, these growth rates
are expected to be valid (in order of magnitude) for $V_A k_z
\lesssim \kappa$ due to our expansion scheme.

Note finally that a similar analysis can be performed for the
inertial modes, but is not very informative; as they appear to be always stable.

\subsubsection{High $\beta$ limit:}

Although we did not investigate this case in much detail, it is
apparent from Eq.~(\ref{disp}) that when $2\Omega S = V_A^2 k_z^2$
[cancellation of the last term in Eq.~(\ref{disp})],
$\omega_\eta=0$ is one of the solutions to the dispersion
relation. At the light of our preceding analyzes, and because this
equality embodies the MRI stability limit in the ideal case, as
recalled above, it is apparent that this relation is the relevant
limit in a small dissipation context as well, generalizing the
result found for $Pm=1$.

\subsubsection{Heuristic explanation:}

\begin{figure}
      \centering
      \includegraphics[scale=0.4]{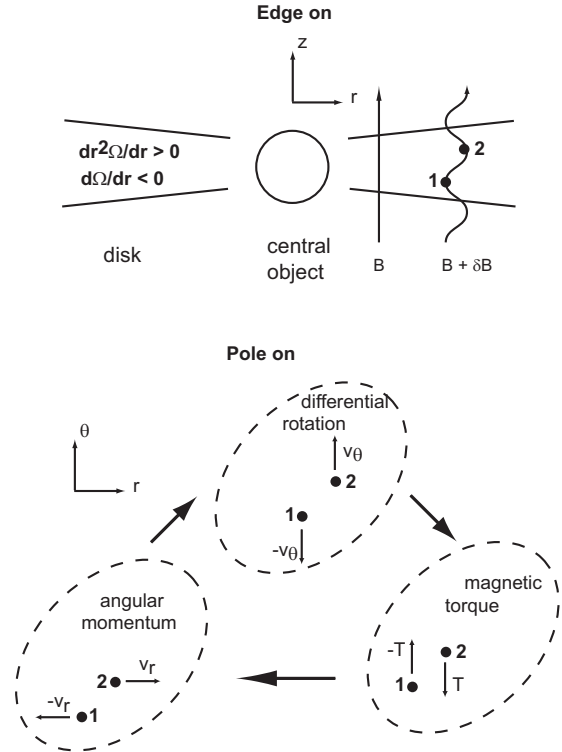}
      \caption{Sketch of the MRI mechanism (see text).\label{mrisketch}}
\end{figure}

To explain the behavior brought to light in Eqs.~(\ref{Rm}) and
(\ref{RmRe}), it is useful to recall the physical origin of the
instability, as discussed, e.g., in \cite{BH03}, in the
dissipation-free limit; the process is sketched on
Fig.~\ref{mrisketch}, for convenience. Assume for definiteness
that one starts by distorting the equilibrium velocity field in
the radial direction with a sinusoidal perturbation in the
vertical direction: $v_y=v_{y_0}\exp\big(-ik z\big)$. The magnetic
field being frozen in the fluid will also develop a radial
component [first term in the right-hand side member of the
linearized induction equation, Eq.~(\ref{linb})]; the shear will
then transform this radial field in an azimuthal one [second term
in the right-hand side member of the linearized induction
equation, Eq.~(\ref{linb})]. The resulting tension force produces
a momentum transfer between fluid particles that have been moved
according to the imposed velocity perturbation [second term in the
right-hand side member of the linearized motion equation,
Eq.~(\ref{linb})]. This force is destabilizing if the angular
velocity decreases with radius: indeed in this case, the inner
particle, moving faster than the outer one, will transfer orbital
momentum to the outer one, thereby reinforcing its inward motion,
an effect mediated by the Coriolis force when seen in the rotating
frame. In this description, marginal stability follows when the
driving mechanism is balanced by the usual tension restoring force
(the piece not connected to the generation of magnetic field from
the mean shear).

What does dissipation change to this picture ? For definiteness,
let us focus on marginal stability and let us consider only
resistive dissipation for the time being (``low" viscosity
limit). In this limit, the magnitude of the velocity and magnetic
fields in the various steps of the instability mechanism described
above are controlled by dissipation processes so that one may
again go through the preceding process step by step, assuming
equilibrium at each step. The magnitude of the radial magnetic
field in this context results from the balance between the motion
driving and field dissipation:

\begin{equation}\label{br}
-ik B_0 v_y = \eta k^2 b_y,
\end{equation}

\noindent while the shearing generation of the azimuthal field
from the radial one is also balanced by resistive dissipation:

\begin{equation}\label{btheta}
  S b_y = \eta k^2 b_x.
\end{equation}

\noindent Both relations follow from the induction equation in the
marginal stability limit, except for the term dropped in
Eq.~(\ref{btheta}), which leads to the usual magnetic tension
stabilization and is of no interest in the limit considered here.
The azimuthal force balance then requires that

\begin{equation}\label{vtheta}
  (2\Omega - S) v_y =  i \frac{kB_0}{\mu_0\rho_0} b_x,
\end{equation}

\noindent i.e., $\omega_\eta^2 \kappa^2=2\Omega S V_A^2 k^2$, once
the two preceding constraints are taken into account (inclusion of
$\omega$ in this line of argument does not change the result). As
noted earlier, this relation directly leads to Eq.~(\ref{Rm}).

If one assumes instead that viscous dissipation exceeds the
Coriolis force in magnitude, then the magnetic tension due to the
generation of azimuthal field from the radial one by the shear
should be balanced by viscous dissipation instead of the Coriolis
force in the two horizontal components of the momentum equation,
leading alternatively to $\omega_\eta^2 \omega_\nu^2=2\Omega S
V_A^2 k^2$, i.e. to Eq.~(\ref{RmRe}).

This also relates to the structure of MRI modes. In the limit of a
very small magnetic tension restoring force, the Alfv\'enic branch
is made of $b_x$ dominated modes. The other components of the
magnetic field and the velocity field are of the order of $V_A k$
compared to $b_x$. Therefore, the growth rate is at first
controlled by the dissipation rate of $b_x$, which is related to
the resistivity [first term of the right hand side of
Eq.~\ref{omega_lowB}]. The interaction of the other fields, which
leads to the MRI, is controlled by a term symmetric in $\nu$ and
$\eta$ [second term of Eq.~\ref{omega_lowB})].

\subsubsection{Generic behavior:}

\begin{figure}
      \centering
      \includegraphics[width=9cm]{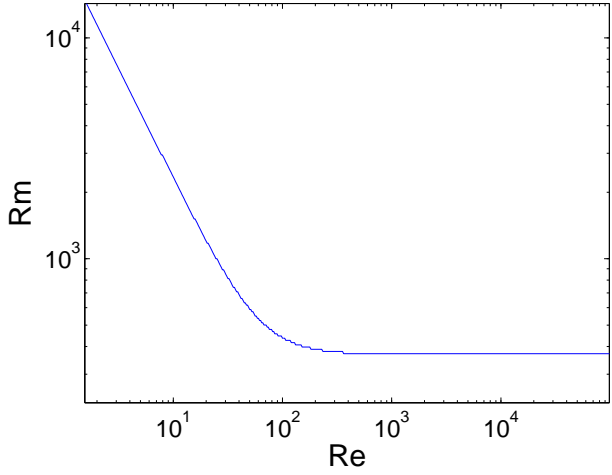}
      \caption{MRI linear stability limit in the $Pm\ne 1$ case
      for $\beta=10^4$.\label{lin_gen1}}
\end{figure}

A more complete view of the stability limits and growth rates
implied by Eq.~(\ref{disp}) may be obtained from exact numerical
solutions for $Pm\ne 1$. Expressing this dispersion relation in
terms of $\omega$ leads to the condition:
\begin{eqnarray}
\nonumber \omega^4&-2ik^2\omega^3(\eta+\nu)-\omega^2\Big(a+k^4(\eta^2+\nu^2+4\eta\nu)+b\Big)\\
\nonumber &+\omega\Big(2ik^6(\eta\nu^2+\nu\eta^2)+aik^2(\nu+\eta)+2ib\eta k^2\Big)\\
\label{disp_full} &+\nu^2\eta^2k^8+a\nu\eta k^4+b\eta^2k^4-c=0,
\end{eqnarray}
with
\begin{eqnarray}
a&=&2k_z^2V_A^2\\
b&=&\kappa^2\gamma^2\\
c&=&k_z^2 V_A^2(2\Omega  \gamma^2 S-k_z^2 V_A^2)
\end{eqnarray}

\begin{figure}
      \centering
      \includegraphics[width=9cm]{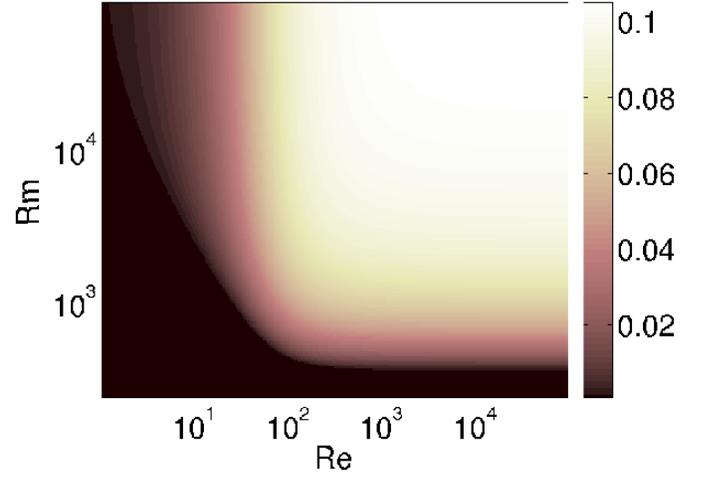}
      \caption{MRI growth rate (arbitrary unit) as a function of viscous and
      resistive dissipation for $\beta=10^4$\label{lin_gen2}}
\end{figure}

To characterize the stability limits as a function of the Reynolds
and the Magnetic Reynolds number ($Rm=SH^2/\eta$), one needs to
choose $\beta$, $\gamma$ and $k_z$. As in the $Pm=1$ case, we take
$k_z=2\pi/H$ and $\gamma=1$ (which are again expected to maximize
the dissipation limits), and solve the relation (\ref{disp_full})
for $\beta=10^4$. The resulting stability limits are shown on
Fig.~\ref{lin_gen1} and the corresponding growth rates on
Fig.~\ref{lin_gen2} (arbitrary units). These results match closely
the analytical limits just discussed: a high $Re$ threshold found
for $Rm\sim 371$, and a low $Re$ threshold found for ${Rm Re}\sim
2.3\times 10^4$, both in agreement with Eqs.~(\ref{Rm}) and
(\ref{RmRe}), respectively. Moreover, significantly lowered growth
rates are observed when $Re\ll 60$ to $80$, as predicted by
Eqs.~(\ref{GRReRm}) and (\ref{Relim}). A similar behavior follows
at much smaller $\beta$. For example, the observed scalings are
identical, and the preceding asymptotic expressions valid within a
factor of 2, down to $\beta$ values of the order of twice the
critical $\beta$ limit.

These results indicate that most of the stability limit behavior
is captured by the approximate relations Eqs.~(\ref{Rm}) and
(\ref{RmRe}) (as well as by the large field $\beta$ limit, where
relevant), whose physical origin has been discussed above.

\subsection{Numerics}

\subsubsection{Equations}

Our objective is to simulate the system of Eqs.~(\ref{base1}) and
(\ref{base2}), with the incompressiblity condition
Eq.~(\ref{base3}), to characterize the dependence of turbulent
transport on the main dimensionless numbers introduced above
($\beta$, $Re$ and $Pm$). We focus on incompressible motions;
indeed, the values of $\alpha$ found in previous investigations
makes us \textit{a priori} expect that compressibility effects
will be small. In any case, this allows us to more effectively
distinguish the effects of the various physical mechanisms at work
in this problem.

First, we simplify the problem from a numerical point of view by
distinguishing the mean laminar shear $\bm{u}=Sy\bm{e_x}$ from the
deviation from this mean $\bm{w}$. The resulting equations read:
\begin{eqnarray}
\nonumber \partial_t \bm{w}+\bm{w}\cdot \bm{\nabla}
\bm{w}&=&-Sy\partial_x \bm{w} - \bm{\nabla}\psi+\frac{\bm{B}\times
(\bm{\nabla}\times \bm{B})}{\mu_0\rho_0}\\
\nonumber  & & +(2\Omega-S)w_y \bm{e}_x-2\Omega w_x\bm{e}_y+\nu\Delta \bm{w}\\
\nonumber\partial_t \bm{B}+\bm{w}\cdot \bm{\nabla} \bm{B}&=&-Sy\partial_x \bm{B}+
\bm{B}\cdot\bm{\nabla} \bm{w}+B_y S \bm{e}_x+\eta \Delta \bm{B}\\
\nonumber \nabla \cdot \bm{w}&=&0\\
\nonumber \nabla \cdot \bm{B}&=&0
\end{eqnarray}

This system is numerically solved using a full 3D spectral code,
using the classical shearing sheet boundary conditions
\citep{HGB95}. This code is now briefly described.

\subsubsection{Numerical code}

The code used for these simulations is an MHD extension of the HD
code used in \citet{LL05}, and extensively described there. This
code is a full 3D spectral (Fourier) code, based on FFTW
libraries, parallelized using the MPI protocol. This kind of code
has many advantages for the simulation of incompressible
turbulence, such as:
\begin{itemize}
\item The incompressibility and solenoid conditions are easily
implemented at machine precision, using a projector function in
Fourier space.
\item The energy budget is much easier to control, leading to a
precise quantification of the energy losses by numerical dissipation.
\item Spatial derivatives are very accurate down to the grid scale
(equivalent to an infinite order finite difference scheme down to
the grid scale).
\end{itemize}

The algorithm used is a classical pseudo spectral method which may
be described as follows. All the derivatives are computed in
Fourier space. However the nonlinear term require special
treatment : in Fourier space, a real space product is a
convolution, for which the computational time evolves as $n^2$,
where $n$ is the number of grid cells. The computation time is
minimized if one goes back to real space, compute the needed
product and then transforms the result to Fourier space. This
procedure (pseudo spectral procedure) is more efficient than a
direct convolution product since the FFT computation time scales
as $n \log n$. However, the finite resolution used in this
procedure generates a numerical artifact commonly known as the
``aliasing" effect (apparition of non physical waves near the
Nyquist Frequency). This effect may be handled through a
dealiazing procedure, in which the nonlinear terms are computed
with a resolution 3/2 higher than the effective resolution used in
the source terms (e.g., \citealt{P02}).

Comparing our spectral code with a ZEUS-type finite difference
code \citep{SN92}, similar results are obtained with a finite
difference resolution two to three times higher than the spectral
resolution. However, FFTs calculations are more computationally
expensive than finite differences, leading to a final
computational time equivalent for both kind of code with the same
``effective" resolution.

\begin{figure}
      \includegraphics[width=0.9\linewidth]{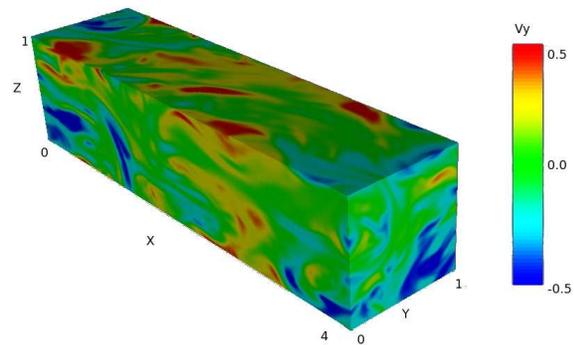}
      \caption{$w_y$ plot (radial velocity) for $\beta=100$, Re=3200, \label{snapshot_mri}}
\end{figure}

All the simulations presented in this paper were performed with an
$xyz$ resolution of $128\times 64 \times 64$ with an aspect ratio
of $4\times 1 \times 1$, $x$ being the azimuthal direction, $y$
the radial direction and $z$ vertical direction. One may change
the physical viscosity and resistivity as well as the magnetic
field intensity ($\beta$). The mean magnetic field (conserved in
the simulations due to the adopted boundary conditions) is aligned
in the $z$ direction. White noise initial perturbations with
respect to the laminar flow are introduced as initial conditions
on all variables. With $\beta=100$, $Pm=1$ and $Re=3200$ one
typically generates flow snapshots as shown on
Fig.~\ref{snapshot_mri} after relaxation of transients; this flow
is quite characteristic of a fully developed 3D turbulent field
\footnote{Movies of some of the simulations
presented in this paper may be found on the web at \\
http://www-laog.obs.ujf-grenoble.fr/public/glesur/index.htm}.

\subsection{MRI behavior near the instability threshold}

The MRI is a weak magnetic field instability, which should be
quenched for $\beta \lesssim 1$ in astrophysical disks. Since the
MRI is assumed to be the source of momentum transport in disks,
and as at least some disks are expected to be close to
equipartition if they are to support magnetically driven ejection
\citep{F97}, on may wonder if this instability is efficient enough
in the vicinity of the strong magnetic field stability threshold.
We investigate this question in an unstratified context here (the
absence of stratification significantly raises the $\beta$
stability threshold).

\begin{figure}
      \includegraphics[width=0.9\linewidth]{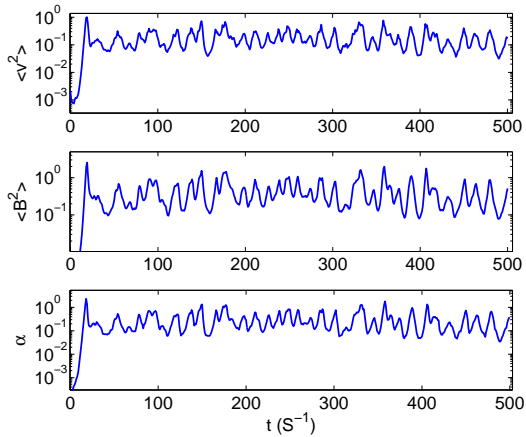}
      \caption{$\beta=100$, Re=3200 run\label{beta=100}}
\end{figure}

\begin{figure}
      \includegraphics[width=0.9\linewidth]{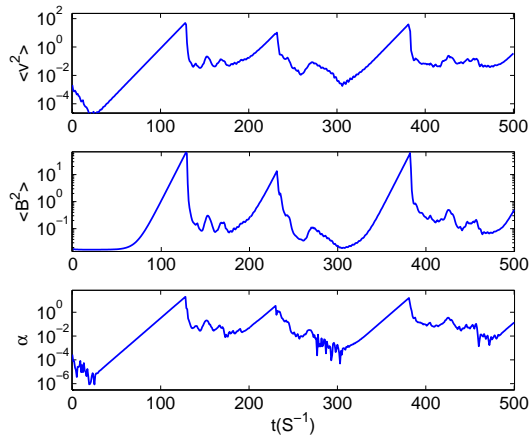}
      \caption{$\beta=30$, Re=3200 run\label{beta=30}}
\end{figure}

We present two simulations. In the first one, $\beta=100$ and
$Re=3200$ (run 1); this run reproduces typical results from the
literature. The second simulation is performed close the $\beta$
threshold, i.e. for $\beta=30$ and $Re=3200$. The time development
of some important quantities is depicted on figs (\ref{beta=100})
and (\ref{beta=30}) for these two runs. One immediately notices a
marked difference between these two simulations. On run 1, we find
a classical MRI behavior, as studied by \citet{HGB95},
characterized by $\alpha\sim 10^{-1}$ and random fluctuations of
all statistical quantities. However, run 2 exhibits strong
exponential growth (``bursts") for about 100 shear times ($\sim$10
orbits), and a sudden drop of fluctuation amplitudes. This
behavior is explained as follows: for such low $\beta$ only the
largest wavelength mode is unstable (and not smaller scales); the
mode amplitude increases for many shear times, as this mode is an
exact nonlinear solution to the incompressible equations of
motions \citep{GX94}. We therefore observe the growth of the
channel flow as seen by \citet{HB92}. However, as this channel
solution reaches sufficiently large amplitude, secondary
instabilities such as the Kelvin-Helmoltz instability quickly set
in and destroy these ordered motions, and a new cycle starts (see
\citealt{GX94} for a detailed description of these secondary
instabilities).

Note that this kind of behavior and related explanation does also apply to
the low Reynolds threshold, since there the smallest scales are viscously
damped and only the largest ones remain unstable. We did observe
this behavior close the low Reynolds threshold, as did
\citet{FSH00} but in an indirect way (see Figs. 2 and 4 of their
paper ), and one can conclude that these bursts are characteristic
of a marginally unstable MRI. Such bursts may be astrophysically
relevant. Indeed, one may wonder about the MRI behavior close to the
dead zone \citep{G96}, where presumably the magnetic Reynolds
number is small, and the instability quenched. If these bursts
exist in real disks, they may quickly destroy this dead zone under
the effects of the strong turbulent motions observed in our
simulations.

\begin{figure}
      \includegraphics[width=0.9\linewidth]{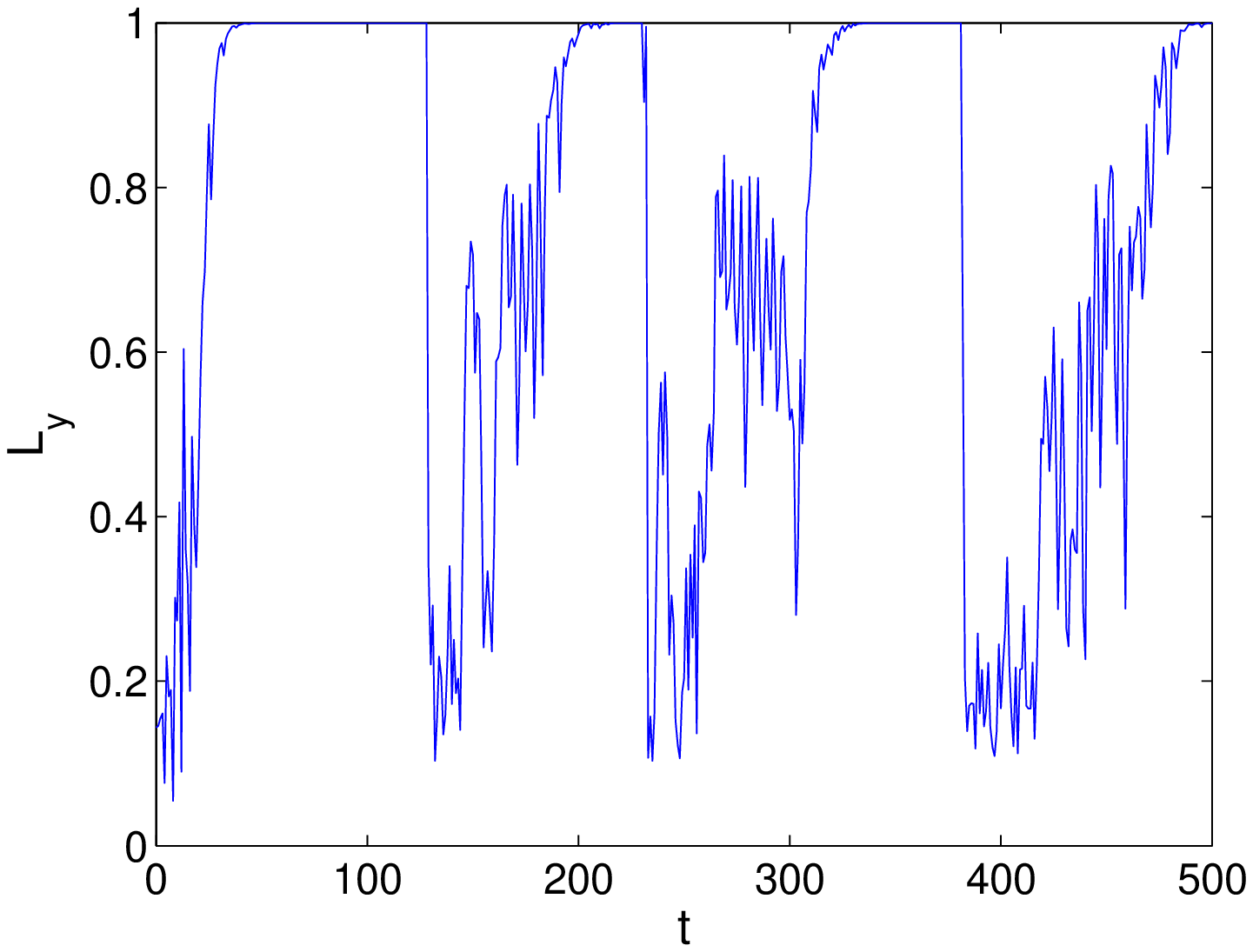}
      \caption{Correlation length of $w_x$ in the $y$ direction
      as a function of time, $\beta=30$.\label{uxlx}}
\end{figure}

\begin{figure}
      \includegraphics[width=0.9\linewidth]{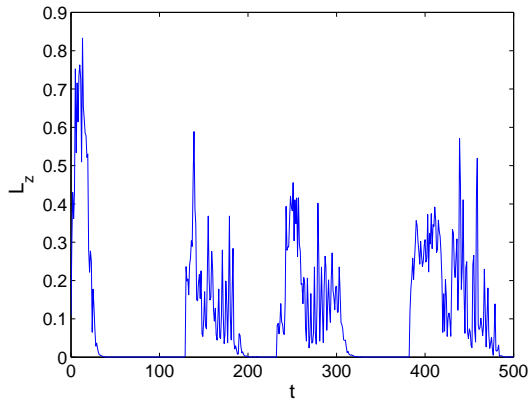}
      \caption{Correlation length of $w_x$ in the $z$ direction as
      a function of time, $\beta=30$.\label{uxlz}}
\end{figure}

Let us have a closer look on these bursts with the help of
correlation lengths defined as

\begin{equation}
L_i=\frac{\int dy_i \int f(x_i) f(x_i-y_i) dx_i}{\int f^2(x_i) \,dx_i}
\end{equation}

\noindent where $i=1,2,3$ is the direction of correlation and $f$
refers either to the velocity or magnetic field. Note that with
this definition, the correlation length vanishes in the $z$
direction for a pure sinusoidal signal, and equals $1$ in the $y$
direction for a channel flow, as a consequence of the shearing
sheet boundary conditions. Therefore, these correlation lengths
provide us with a convenient tool to trace the presence of the
channel flow solution in our simulations. We show on
fig.~\ref{uxlx} and fig.~\ref{uxlz} the evolution of the
correlation length in the $y$ and $z$ direction for the $w_x$
field (a similar behavior is obtained with the other field
components). The behavior of correlation lengths closely follows
what can be seen by monitoring the energy in the deviations from
the laminar flow (fig.~\ref{beta=30}), and indicate the presence
of two main regimes in this simulation. The first regime
corresponds to an exponential growth (``burst") of the channel
flow, for which $L_y$ is found to be equal to the box size and
$L_z=0$ (a careful examination shows that $L_z$ is exponentially
decaying down to $10^{-10}$), indicating the presence of a purely
sinusoidal mode in the $z$ direction in the burst stage. The
second regime is a more classical state for 3D turbulent motion,
with $L_y \simeq 0.5$ and $L_z\simeq 0.4$. Note that $L_y$ grows
on very short time-scales, leading eventually to a new burst
stage.

These correlation lengths disclose numerical artifacts in the
first regime. In a real disk, one would expect a loss of
correlation in the radial direction on a scale of the order of a
few scale heights: indeed, the typical frequency involved in these
phenomena is of the order of the Keplerian frequency and a signal
can't propagate faster than the sound speed, leading to a maximum
correlation length of a few scale heights.

Similarly, the vanishingly small vertical correlation length for
the channel flow solution is also an artifact of the adopted
boundary conditions. A more realistic result would follow if one
were to take into account the vertical stratification and set the
boundary conditions far enough from the disk midplane. More
generally, our results are probably not directly applicable to a
real disk, but they shed some light on what the generic behavior
of the MRI would look like near various stability thresholds, even
though different aspect ratio and boundary conditions should be
investigated before firm conclusions can be drawn.


Finally, the behavior exemplified in our simulations suggests that
assuming $\alpha$ constant would poorly represent the transport
behavior close enough to the marginal stability limit.
Time-dependent transport models are needed in such a context. Real
disks may not operate close to the strong field limit unless some
(unknown) back-reaction loop is at work, or unless (more
realistically) the magnetic field varies in a systematic way with
radius throughout the disk; consequently, the bursting behavior
observed here may imply a similar ejection variability in the
relevant regions of jet-driving disks. Note however, that our
``mean" equivalent $\alpha$ is rather large ($\alpha\simeq 5$),
leading to question the role of the ignored fluid compressibility
in these cases; it is quite possible that couplings to
compressible modes may effectively limit the magnitude of the
bursts.

\subsection{Magnetic Prandtl dependence of transport coefficients}

\begin{figure}
      \centering
      \includegraphics[width=0.9\linewidth]{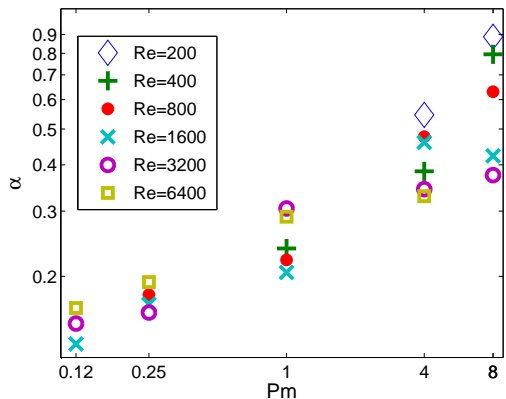}
      \caption{Prandtl effect for $\beta=100$ \label{prandtl} }
\end{figure}

All previously published simulations were performed without
numerical control of the dissipation scales and dissipation
processes. However, as pointed out earlier, such a control is
required to ascertain convergence. In this section, the role of
the Reynolds and Prandtl numbers defined in section \ref{analytic}
is examined. In particular, the Prandtl number allows us to change
the ratio of the viscous and resistive dissipation scales.
Unfortunately, deviations from $Pm=1$ are quite demanding
numerically, since one wants to resolve both the velocity and
magnetic dissipations scales. We present on Fig.~\ref{prandtl} the
result of such simulations: we plot the mean transport coefficient
($\alpha$) as a function of the Prandtl number, for various
Reynolds numbers (the Reynolds number quantifies the viscous
dissipation scale). Statistical averages are computed over 500
shear times, and start after the first 100 shear times to avoid
pollution by relaxation of the initial transient dynamics. From
these plots, one finds a significant correlation between the
Prandtl number and the transport coefficient, leading to

\begin{equation}\label{Pmalpha}
\alpha\propto Pm^\delta \qquad \mathrm{for} \left\{ \begin{array}{c}
0.12<Pm<8\\
200<Re<6400
\end{array} \right.
,
\end{equation}

\begin{figure}
      \centering
      \includegraphics[width=0.9\linewidth]{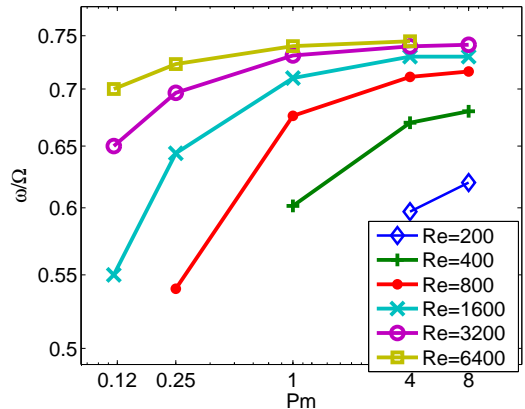}
      \caption{Linear growth rate of the largest mode for various ($Re,Pm$) at
$\beta=100$ \label{GR} }
\end{figure}

\noindent with $\delta$ in the range $0.25$ --- $0.5$. Note that this
results show that the transport coefficient depends on $Re$ and $Rm$
 via $Pm$, at least in the $Pm$ range concidered in this paper. This may be
seen on Fig.~\ref{prandtl} as a small vertical dispersion (variation of both
$Re$ and $Rm$ at constant $Pm$) compared to the effect of a single $Pm$ change.
Although
this section is the briefest of the paper, this result constitutes
the most important finding of this investigation (and the most
computationally intensive one!).

Note that the numerical results obtained at very high Reynolds
number and high Prandtl number are only marginally resolved,
mainly because of a very short magnetic dissipation scale. This
remark may explain that the two points at $Pm=8$ lie somewhat
below the mean of the other results. Our preliminary tests at
higher resolution seem to show that a higher transport obtains at
higher resolution at $Pm=8$ and $Re=6400$, which confirms a limit
due to resolution in these high $Pm$ runs. This behavior is easily
understood, since the finite numerical resolution enforces a
numerical dissipation scale (roughly equal to the grid scale),
which is obviously the same for the magnetic and velocity fields.
Therefore, at high $Pm$, the effective magnetic dissipation scale
is forced to be larger than the expected one, leading to an
altered spectral distribution and a smaller ``numerical Prandtl".

One may wonder if this Prandtl dependence may be correlated to the
linear growth rates discussed before. To this effect, we plot the
linear growth rate of the largest mode for the different
simulations used for this study on Fig.~\ref{GR}. Similar results
follow when replacing the growth rate of the largest mode by the
maximum growth rate. Although the idea of a transport efficiency
controlled by the linear growth rate is widely spread in the
Astrophysical community, this plot shows us that, at least for
this example, the linear growth rate doesn't explain the transport
behavior observed on Fig.~\ref{prandtl}. Moreover, it appears
that, as one may suspect from Eq.~(\ref{omega_lowB}), the growth
rate is not controlled only by $Pm$, but also by some complex
combination of $Re$ and $Rm$. \cite{UMR07} tried to get this kind
of $\alpha-Pm$ correlation analytically, using a weakly non linear
analysis of the channel flow. This study leads to a stronger
$\alpha-Pm$ correlation with $\delta=1$ in the limit $Pm\ll1$,
which appears to be quite different from our full 3D numerical
results. Note however that their analysis belongs to very different
boundary conditions (rigid instead of shearing sheet) and there results
are therefore not of direct relevance to our numerical investigation;
nevertheless both studies point out the role of the Prandtl number.
In any case, one needs to find some full nonlinear theory
to explain the transport dependance on $Pm$.

The observed correlation indicates the existence of a
back-reaction of the small magnetic field scales on the large ones
(at least for the range of Reynolds and Prandtl numbers explored
here). Note that this effect is expected to saturate at some
point, since in the limit $Pm\rightarrow 0$ with $Re\rightarrow
\infty$ and $Rm$ kept constant, Eq.~(\ref{Pmalpha}) predicts a
vanishing transport in spite of the existence of the linear
instability. This issue is further discussed in the conclusion,
the Reynolds number limitation of our investigation being the most
serious here. In
any case, the exact implications of these findings remain to be
understood, but may potentially be quite important since the
Prandtl number varies by many orders of magnitude in astrophysical
objects. For example, \cite{BS05} suggest that values as small as
$Pm\sim 10^{-8}$ might be found in young stellar objects, while
$Pm\sim 10^4$ would be more typical of AGN disks. These estimates
are highly uncertain; even a substantially narrower range is of
course out of reach of present day computers.

Finally, this kind of back-reaction points out the potential role
of small scale physics (dissipation scales) on the properties of
turbulence at the largest available scales (disk height scale).
This argues for a careful treatment of the role of dissipation and
reconnection processes on the turbulence transport
characterization.

\section{Discussion}

In this paper, we have investigated the role of local
dimensionless numbers on the efficiency of the dimensionless
turbulent transport. To this effect, we have first generalized
previously published linear stability limits, to account for the
presence of both viscous and resistive dissipation. Namely, we
have confirmed in all cases that the large field marginal
stability limit is characterized by a constant plasma $\beta$
parameter, of order 30 in the shearing sheet unstratified context
(but more likely of order unity in real, stratified disks). When
marginal stability follows from dissipation and not magnetic
tension stabilization, we have found that the marginal stability
limit is captured by two asymptotic regime: a large Reynolds
($Re$), small magnetic Reynolds one ($Rm$), with a marginal
stability limit $Rm\sim \beta^{1/2}$, and a small Reynolds, large
magnetic Reynolds number one, where $ReRm\sim 10^2 \beta^{1/2}$. A
phenomenological explanation has been provided for this behavior.

In the previous section, we have investigated the behavior of the MRI
near the low $\beta$ instability threshold; in our simulations,
$\beta=30$, a value representative of the large field threshold in
our simulation box. In vertically stratified disks, this threshold
obtains for much smaller values, typically $\beta \sim 1$ (Gammie
\& Balbus 1994). We found, somewhat unexpectedly, that turbulent
transport is significantly enhanced through burst events, even
surprisingly close to the marginal stability threshold. As pointed
out earlier, this behavior is physical and not numerical. The use
of periodic boundary conditions (vertical) or semi periodic
(radial) boundary conditions may enhance the role of the channel
flow solution which is responsibly for this behavior, and a real
disk channel flow may break up sooner than observed in our local
simulations, leading to smaller burst magnitudes. Moreover,
$\alpha>1$ leads to supersonic motions and compressible numerical
simulations are needed to properly quantify the phenomenon, which
may exhibit new secondary compressible instabilities in such a
context. All these issues lead to the conclusion that low $\beta$
MRI would produce weaker bursts and therefore smaller transport
coefficient than observed in our simulation. However, there is no
physical reason why the turbulence bursts would be suppressed, and
we believe that these bursts may be a strong signature of regions
of stratified disks where MRI-driven turbulence is driven close to
the marginal stability threshold.

The most important new result reported in this paper is a
correlation between the transport efficiency, and the magnetic
Prandtl number, leading to a higher transport coefficient for
larger Prandtl numbers. As in the case of the bursting behavior
discussed above, the boundary conditions used in these simulations
play some role in the result. However, the possible biases are
less obvious and tests with plane radial walls need to be
performed to get a grasp on boundary condition effects. Moreover,
one needs to check the correlation at higher resolutions, and if
possible higher Prandtl numbers, using different kind of codes to
get a better characterization and a physical description of the
phenomena involved in this observation.

More specifically, a puzzling fact points towards a potential bias
due to the shearing sheet boundary conditions. In non-magnetized
shear flows, transport in the subcritical regime, far enough from
the marginal stability limit scales like $1/Rg$ where $Rg$ is the
subcritical transition Reynolds number \citep{LL05}. Closer to the
marginal stability limit, and in the supercritical regime (e.g.,
when the Rayleigh stability criterion is not satisfied), transport
is enhanced with respect to this scaling, but one always has
$\alpha < 1/Rc$ where $Rc$ is the critical Reynolds number of
linear instability. However, for MRI-driven turbulence, one has
$\alpha >$ or$\gg 1/Rc$, as can be checked from our results. Close
to the marginal stability limit, this enhanced efficiency is
related to the existence of the channel flow solution, as
discussed above. As each linear mode is a nonlinear solution to
the incompressible problem, one may ask whether this enhanced
transport, which is observed also far from the marginal stability
limit, is not an artifact of the shearing sheet boundary
condition, which allows such nonlinear coherent modes to develop.
This behavior is not necessarily unphysical or irrelevant to
actual disk systems, but this point needs to be checked in the
future.

Finally, let us come back to the magnetic Prandtl number
dependence of $\alpha$. As pointed out earlier, the dependence of
the transport efficiency on the magnetic Prandtl number indicates
a back-reaction of small scales on large ones. We make here a few
comments on this feature. In particular, we shall argue that this
behavior must saturate at low and large enough $Pm$. The magnetic
Prandtl number is related to the ratio of the viscous $l_\nu$ and
resistive $l_\eta$ dissipation scales, the exact relation
depending on the shape of the turbulent energy spectrum. Generally
speaking, the Prandtl number varies monotonically with the ratio
$l_\nu/l_\eta$, and one expects $Pm \ll 1$ (resp. $Pm \gg 1$) when
$l_\nu/l_\eta \ll 1$ (resp. $l_\nu/l_\eta \gg 1$). The spectrum of
the largest scales tends to be flatter than usual turbulent
spectra due to the role of the linear instability, down to the
scale where the magnetic tension prevents the instability to occur
(most probably, this ``instability sector" of the spectrum only
represents a small part of the overall turbulent spectra of actual
disks, because of their enormous Reynolds numbers). Leaving aside
these largest scales, for $Pm \ll 1$, the spectrum is expected to
be Kolmogorovian and anisotropic down to the resistive dissipation
scale \citep{GS95}, while below this scale and down to the viscous
scale, the velocity spectrum is the usual Kolmogorov velocity
spectrum and the magnetic spectrum drops much faster. On the other
hand, for $Pm\gg 1$, the spectrum should be Kolmogorovian down to
the viscous dissipation scale \citep{GS95}, while the magnetic
spectrum should scale like $k^{-1}$ below the viscous dissipation
scale and down to resistive scale \citep{CLV03}. It is therefore
tempting to see in a difference of accumulation of magnetic energy
at small scales the cause of the back-reaction of these scales to
the largest ones, which would create the observed magnetic Prandtl
number dependence of the turbulent transport efficiency.
Nevertheless, in both small and large Prandtl number settings,
turbulent motions in the inertial range are random in phase, so
that one expects that to lowest order, coupling of the turbulent
spectrum with the largest MRI unstable scales vanishes. To next
order, the steepness of the Kolmogorov spectrum indicates that the
strength of the coupling decreases with increasing Reynolds number
in the vicinity of the viscous dissipation scale, suggesting that
at large enough Reynolds numbers, the Prandtl dependence should
saturate (a potential caveat to this argument being the possible role
played by a small scale field generation through dynamo action). Such a
saturation was not observed in our simulations,
although a weak dependence of our results on the magnitude of the
Reynolds number may be detected on Fig.~\ref{prandtl}; however,
such an effect might also arise from resolution requirements,
which makes our lower Reynolds number results confined to the
larger Prandtl number range. Unfortunately, our results can hardly
be improved upon with the present generation of computers, leaving
the question of the Reynolds number saturation of the Prandtl
number dependence open, as well as the overall difference in
transport efficiency between the small and large Prandtl number
cases. Resolving this issue is crucial to ascertain the role of
the magneto-rotational instability in disk transport.

\section*{Acknowledgements}

The simulations presented in this paper has been performed both at
IDRIS (French national computational center) and at the SCCI
(Grenoble Observatory computational center). The authors
acknowledge fruitful discussions on the issues discussed, with
Steve Balbus, S\'ebastien Fromang, Gordon Ogilvie, and John
Papaloizou.

%%-----------------------------
%%      your bibliography
%%-----------------------------
%In preparing the reference list please adhere to the following format.
% Attention should be paid to the order of the items in each reference
% and to the punctuation used. Please see Sect. 4 in the User's Guide
% that comes with the new macro package.

%Bohr, N., Einstein, A., \& Fermi, E. 1992, MNRAS, 301, 257 (BEF)
% Curie, M., \& Curie, P. 1991, A\&A, 248, 612
% de Gaulle, C. 1996, Solar Phys. (Oxford: Oxford Univ. Press)
% Heisenberg, W., \& West, C. N. 1993, Australian J. Phys., 537, 36  (Paper III)
% Laurel, S., \& Hardy, O. 1994, Active Galactic Nuclei, in The Evolution
% and Distribution of Galaxies, ed. W. Churchill, F. D. Roosevelt, \& J.
% Stalin (New York: Wiley), 210

\bibliographystyle{mn2e}

\bibliography{glesur}

%\bibitem{01}Balbus, S.A., \& Hawley, J.F. 1991, ApJ, 376, 214
%\bibitem{0b}Balbus, S.A., Hawley, J.F., \& Stone, J.M. 1996, ApJ, 467, 76
%\bibitem{02}Fleming, T.P., Stone, J.M., \& Hawley, J.F. 2000, ApJ, 530, 464
%\bibitem{03}Gammie, F.G., \& Balbus, S.A. 1994, MNRAS, 270, 138
%\bibitem{04}Goodman, J., \& Xu, G. 1994, ApJ, 432, 213
%\bibitem{05}Hawley, J.F. 2000, ApJ, 528, 462
%\bibitem{06}Hawley, J.F., \& Balbus, S.A. 1992, ApJ, 400, 595
%\bibitem{07}Hawley, J.F., Balbus, S.A., \& Winters, W.F. 1999, ApJ, 518, 394
%\bibitem{08}Hawley, J.F., Gammie, C. F., \& Balbus, S.A. 1995, ApJ, 440, 742
%\bibitem{09}Lesur, G., \& Longaretti, P-Y. 2005, A\&A, 444, 25
%\bibitem{10}Lesur, G., \& Longaretti, P-Y. 2007 \emph{in prep.}
%\bibitem{11}Richard, D., \& Zahn, J-P. 1999, A\&A, 347, 734
%\bibitem{12}Shakura, N.I., \& Sunyaev, S.A. 1973, A\&A, 24, 337
%\bibitem{13}Stone, J.M., Hawley, J.F., Gammie, C.F., \& Balbus, S.A. 1996, ApJ, 463, 656

\end{document}